\documentstyle[12pt]{article}
\newcommand{\al}{\alpha}             \newcommand{\bt}{\beta}
			
\newcommand{\ga}{\gamma}

\newcommand{\pa}{\partial}

\newcommand{\be}{\begin{equation}}	
		
\newcommand{\ee}{\end{equation}}  \newcommand{\na}{\nabla}
\newcommand{\bq}{\begin{eqnarray}} \newcommand{\r}{\rho}
\newcommand{\eq}{\end{eqnarray}}   \newcommand{\ar}{\rightarrow}

\begin{document}
\baselineskip= 24 truept
\begin{titlepage}
\title { Singular Limits and String Solutions}
\author{\sc S. Pratik Khastgir and Alok Kumar  \\
\\
 Institute of Physics, Bhubaneswar-751 005, India. \\
 email: pratik@iopb.ernet.in, kumar@iopb.ernet.in }
\date{}
\maketitle
\thispagestyle{empty}
\vskip .6in
\begin{abstract}
\vskip .2in
It is shown that new leading ($\al'$) as well as all-order
solutions of String theory can be obtained by taking
appropriate singular limits of the known solutions.
We give several
leading order solutions for the bosonic as well as
the heterotic string.
We then present all-order forms of the previously known two
dimensional cosmological solutions.
An all-order form for the cosmological solution
in three dimensions is also predicted.
The physical implications of our results are discussed.
\end{abstract}
\bigskip
\flushright {IP/BBSR/93-72}
\vfil
\end{titlepage}
\eject

Classical solutions of string theory and their physical
implications have been explored extensively recently.
The search for the solutions
of string theory has been done mainly along two lines.
First, several leading
order ($\al'$) solutions have been obtained by solving the
equations of
motion of the string effective action or the vanishing of the
one-loop $\bt$-function equations\cite{call,mandal}.
However this has the drawback
of ignoring important stringy effects. The exact string solutions in
the curved background, on the other hand, are derived from the
Lagrangians of several exact conformal field theories, such as WZW
and coset conformal field theories\cite{witt,ish}.

In this paper we show that appropriate singular limits of the above
backgrounds, both exact as well as effective  (leading order
solution), give rise to new consistent solutions of string theory.
The backgrounds generated in this way
are not equivalent to the original ones. For example, by writing the
two dimensional black hole in an appropriate coordinate system,
the two dimenional cosmological solution\cite{gmv} and  the
Liouville theory can be obtained by taking two different limits.
Inspired by this result, we take singular
limits in various other physical circumstances and obtain new
solutions.  A set of three dimensional
cosmological backgrounds are obtained in this way from the
3-D black string\cite{horn}.
Similarly, a set of new solutions are derived
from the two dimensional charged black hole of the heterotic
string theory\cite{nappi}. These solutions are
characterized by the vanishing of the corresponding cosmological
constant.

One can
also extract all-order  solutions of string theory in this manner.
In this paper we have given the all-order extension of the
2-D cosmological solution\cite{gmv}. The singular limit
of the exact three dimensional black string\cite{sfet} also
generates cosmological backgrounds
which we conjecture to be exact to all orders.

We believe that our results can also be used in further
classification of
the solutions of string theory. Earlier, it has been shown
that a number of such solutions can be classified by the $O(d, d)$
symmetry transformations\cite{mv}. However these
transformations correspond,
in the language of the conformal field theory (CFT), to the
deformations by the marginal operators \cite{hassan} and therefore
leave the cosmological constant, a
measure of the central charge, unchanged. But as we mentioned above, the
solutions generated by singular limits correspond to zero
cosmological constant. As a result we may be able to show
new connections among the classical vacua of string theory.
A possible CFT interpretation of our results
probably requires deformations by relevant
operators.

Now, in table-1 below we present our results for the leading order
solutions. We use the notations and conventions of
\cite{cal2}. In column-1, we give the original
solutions. The final solutions are presented
in column-2.

We now discuss the main points about the solutions presented in
table-1. The first column for the
solution (I) represents the dual of the 2-D black hole of
ref.\cite{witt}. The corresponding cosmological
constant is $-4b^2$. For this case, there are in fact more
than one ways to reach to the solution of column-2. First,
one can directly take the limit $A\ar 0$, $b\ar 0$ such that ${A\over
b^2} =1$, redefine $r\ar t,\; t\ar x$ and
reach to the final solution with zero cosmological constant.
Alternatively, by a coordinate transformation, one can
rewrite the 2-D black hole
background of column-1 in the form:
\be
	ds^2 = dt^2 + {1\over {(at + bx)}^2}dx^2 \qquad{\rm and}\qquad
	\phi = -\log{(at + bx)} + {\rm const.},	\label{bh}
\ee
which for $b\ar 0$ reproduces the
solution of column-2. From eqn.(\ref{bh}), we also see that
in another limit $a\ar0$ one obtains the
Liouville theory together with an extra free string coordinate.
The final solution of column-2 was discussed in ref.\cite{gmv}
and describes
the string moving in a  superinflationary universe for $t < 0$.
Later on we will present an extension of this solution to all-orders
in $\al'$.

\vfil
\eject

{\center{{\bf Table 1}: Leading order solutions}}
\begin{center}
\begin{tabular}{|l|lc|} \hline
\multicolumn{1}{|c|}{Original solution} &
\multicolumn{2}{|c|}{Final solution}
\\
\hline
     $g_{rr}=1$   & $g_{tt}=1$&\\
     $g_{tt}=A\coth^2br$ & $g_{xx}={1\over t^2}$& I\\
     $\phi=-{1\over 2}\log\sinh^2 br+{\rm const.}$  &
	 $\phi'=-\log t+{\rm const.}$&\\
\hline
     $g_{rr}=1$   & $g_{tt}=1$&\\
     $g_{tt}=A\tanh^2br$ & $g_{xx}=t^2$& II\\
     $\phi=-{1\over 2}\log\cosh^2 br+{\rm const.}$  &
	 $\phi={\rm const.}$&\\
\hline
    $g_{tt}=-(1-{M\over
r})$  & $g_{tt}=(1+{q^2\over t^2})^{-1}$&\\
     $g_{xx}=(1-{Q^2\over {Mr}})$ & $g_{xx}=(1
	 +{q^2\over t^2})$&\\
	 $g_{rr}=(1-{M\over r})^{-1}(1-{Q^2\over {Mr}})^{-1}
	 {k\over {8r^2}}$  &
	 $g_{yy}=-{1\over t^2}$& III.1\\
     $B_{tx}=-{Q\over r}$   &
	 $B_{xy}=-{q\over t^2}$&\\
     $\phi=-{1\over 2}\log r+{\rm const.}$  & $\phi=
	 -\log t+{\rm const.}$&\\
\hline
      & $g_{tt}=1$&\\
      & $g_{xx}={1\over t}$&\\
     Solution III.1  &  $g_{yy}=-{1\over t}$& III.2\\
     & $B_{xy}=-{1\over t}$&\\
     & $\phi=-{1\over 2}\log t
	 +{\rm const.}$&\\
\hline
    $g_{tt}=- {(1 - {2M\over r} + {Q^2\over r^2})}$   &
	 $g_{tt}= {(1+{q^2\over t^2})}^{-1}$ &\\
     $g_{rr}={1\over Q^2 r^2}
			 {(1 - {2M\over r} + {Q^2\over r^2})}$ &
	$g_{xx}= -{1\over t^2} {(1+{q^2\over t^2})}$		 & IV\\
     $	F_{t r} = {{\sqrt 2}Q\over r^2}$ &  $
	 	F_{t x} = - {{2\sqrt 2} q\over t^3}$&\\
     $\phi=-{1\over 2} \log r+{\rm const.}$   & $\phi
	 =-\log t+{\rm const.}$&\\
\hline
\end{tabular}
\end{center}

\bigskip
{\center{{\bf Table 2}: All-Order Solutions}}
\begin{center}
\begin{tabular}{|l|lc|} \hline
\multicolumn{1}{|c|}{Original solution} &
\multicolumn{2}{|c|}{Final solution}
 \\
\hline
     $g_{rr}=1$   & $g_{tt}=1$&\\
     $g_{tt}= A \coth^2 br
	(1- {2\al'b^2\over {\sinh^2 br}})^{-1}$ & $g_{xx}
	={1\over {(t^2- 2\al')}}$& V\\
     $\phi=- {1\over 4} \log (\sinh^4 br $   & $\phi
	 =- {1\over 4} \log (t^4 - 2\al't^2)$&\\
	$~~~~ - 2\al' \sinh^2 br)
	+{\rm const.}$   & $~~~~	+{\rm const.}$&\\
\hline
     $g_{tt}=-(1-{r_+\over r})$    & $g_{tt}={(1 + {{q^2
	 + 2\al'}\over t^2})}^{-1}$&\\
     $g_{xx}= (1-{{r_--r_q}\over {r-r_q}})$  & $g_{xx}
	={(1 + {q^2 \over {t^2+ 2\al'}})}$& \\
	     $g_{rr}=- {k'\over 8r^2} {(1-{r_+\over r})}^{-1}
	 {(1-{r_-\over r})}^{-1}$
	 & $g_{yy} = -{1\over t^2}$&VI\\
	$B_{tx}={\sqrt{{(r_-r_q)}\over r_+}}{{r-r_+}\over {r-r_q}}$
	 & $B_{xy}=-{q\over {t^2+2\al'}}$&\\
    $\phi=-{1\over 4}\log(r(r-r_q))-{\rm const.}$
	 & $ \phi=-{1\over 4} \log (t^4 + 2 \al' t^2)$&\\
\hline
\end{tabular}
\end{center}
\vfil
\eject

The next solution (II) in table-1 is dual\cite{mv,gmv}
to (I) and describes the
string dynamics in {\it Milne} space-time. The metric in (II) can in
fact be transformed to a Minkowski metric by a general coordinate
transformation\cite{gmv}. Therefore it is already an
exact solution of string theory to all-orders.

The background written in the first column of solution (III.1)
is the charged black string of ref.\cite{horn}. Here we
treat $k$ as a parameter in the solution of the effective action
which gives the cosmological constant $\Lambda = -{8\over k}$.
By a coordinate transformation $r= -M \sinh^2 b\r$, and scaling
$t\ar {\sqrt A} t$ it can be written as
\be
	ds^2 = - A \coth^2 b\r dt^2 + {(1 + {Q^2\over M^2 \sinh^2 b\r})}dx^2
			+ {b^2 k \over 2}
			{(1 + {Q^2\over M^2 \sinh^2 b\r})}^{-1} d\r^2,
\ee
\be
	B_{tx} =  {{Q\sqrt A}\over {M \sinh^2 b\r}},
	\qquad {\rm and} \qquad
	\phi   = -{1\over 2}\log \sinh^2 b \r + {\rm const.}.
\ee
Then by renaming the coordinates as,
$t \ar y$, $\rho \ar t$, we get in the
limits, $b \ar 0$, $A\ar 0$, $k\ar \infty$, ${Q\over M}\ar 0$
with ${Q\over {M b}} = q$, ${A\over b^2} = 1$, and $b^2k =1$,
the background (III.1).
It gives the cosmological evolution in the
presence of a nontrivial antisymmetric tensor.
We have verified that they satisfy
the one-loop beta function equations and are basically of the
type given in ref.\cite{gmv}. We will later on present an all-order
extension of this background also.

The above three examples conclusively
show that the limiting procedure outlined in
the introduction can generate new solutions belonging to a different
class than the original ones. We now present its usefulness by
generating several new solutions.
The solution (III.2)  has been generated from (III.1)
by a coordinate
transformation $t \ar q \sinh b{\sqrt t}$, and by appropriate
limits and redefinitions. Once again it
describes the cosmological evolution in the presence
of an antisymmetric tensor field.
But unlike (III.1), the charge of the antisymmetric tensor is fixed to
a definite value. The singular limits taken in this case do
not leave any free parameter.

The solution (IV) has been generated from the charged black hole of
the heterotic string\cite{nappi} in the same manner as
outlined in detail for the 3-D black string.
This now describes the cosmology in two dimensions
in the presence of a nontrivial gauge field. We have again
verified that these backgrounds satisfy the one-loop
beta function equations for the 2-D heterotic strings. In the limit
$q\ar0$ this solution reduces to (I).
By a coordinate transformation, $T^2 = t^2 + q^2$,
the curvature scalar is given by $ R = -4 [ {(T^2 + 3 q^2)}/{(T^2 -
q^2)}^2]$. The singularities are therefore at the points
$ T = \pm q$.  The metric describes an
expanding space-time in the regions $-\infty <T < -q$ and
$0 < T < q$ and a contracting one  in the regions
$ \infty > T > q$ and $ 0 > T > -q$.

We now come to the discussions of the exact solutions presented in
table-2. The first one (V) is the all-order extension
of the cosmological solution (I). It is obtained
from the exact black hole solution presented in \cite{dvv,tset}
by taking the same limit as for the solution (I).
We have by now already shown, through a number of examples, that singular
limits give consistent string solutions.
Therefore we expect that the field equations satisfied by the exact
black hole will also be satisfied by the new solution (V).
In another context in string
theory, it has been pointed out that singular limits, known as the
In\"onu-Wigner contractions\cite{majum}, generate
consistent conformal algebra starting form a known one.

We have explicitly verified that the backgrounds
(V) satisfy the beta
function equations upto two-loops\cite{tset,jones,ntset}:
\bq
	R_{\mu \nu} + 2 \na_{\mu} \na_{\nu}\phi
	+ {\al'\over 2} R_{\mu \al\bt\ga} R_{\nu}^{~\al\bt\ga}& = 0  \cr
	{(\na \phi)}^2 - {1\over 2} \na^2 \phi +
	{\al'\over 16} R_{\mu\nu\al\bt} R^{\mu\nu\al\bt}& = 0
\eq
At higher loop orders, the beta functions are not
uniquely defined\cite{tset,jones}.
They depend on the renormalization scheme chosen to
regularize the theory. But we do expect that there exists a
scheme in which
the backgrounds (V) satisfy the beta function equations to all-orders.
As a further check, we observe that the backgrounds (V) are related to
the leading order solution (I) by a field redefinition:
\bq
	g'_{\mu\nu} & = g_{\mu\nu} - 2 \al'\pa_{\mu}\phi \pa_{\nu}\phi
					{1\over {1+{\al'\over 2}R}}
					+ 2 \al' g_{\mu\nu} {(\pa \phi)}^2
					{1\over {1+{\al'\over 2}R}}		\cr
 	& \phi' = \phi - {1\over 4} \log [1 + {{\al'}\over 2} R]
										\label{redef}
\eq
These are the same redefinitions given in ref.\cite{ntset}
which connect the exact and the leading order black
hole solutions.

The curvature scalar for the solution (V) is given by
$ R = -4 [ {(t^2 +  \al')}/{(t^2 -2\al')}^2]$ and the
Hubble parameter\cite{ven}
is given by $H = - [t/{(t^2-2\al')}]$. The curvature singularities
are at the points $ t= \pm {\sqrt 2\al'}$.
One can verify that the
geometric structure of the space-time is similar to the one for
solution (IV).

As we have mentioned above, the solution (II) being flat is
already exact.
This can also be seen by taking the appropriate singular limits in
the exact black hole solution.
Also, since in (II) $R=0$ and $\phi= {\rm const.}$, the redefinitions
(\ref{redef}) now give back the original solution.

The solution (VI) in table-2 has been generated from the exact three
dimensional charged black string presented in
eqns. (7.13) - (7.16) of ref.\cite{sfet}.
It  matches with (III.1) for $\al'=0$.  Alternatively, for $q=0$,
but $\al'\neq 0$, we get the all-order solution (V), together with a
free string coordinate, by the coordinate
transformation, $t^2 \ar t^2 + 2\al'$. We conjecture that (VI)
is in fact  an all-order solution.
Unfortunately, in the presence of a nontrivial $B$ field, the beta
function becomes dependent on the renormalization scheme even at the
two loop level \cite{ptset} and the explicit verification
of the field equations becomes complicated. Alternatively,
to show that it is an all-order solution a field
redefinition of the type given in eqn.(\ref{redef}) has
to be found between the solutions (III.1) and (VI).
We hope to report on this issue in future.

There are other important aspects of the exact solutions (V) and (VI)
which should be examined in detail. For example, it will be
interesting to get an exact conformal field theory for these
backgrounds. Since these solutions correspond to zero cosmological
constant, we can think of two ways to describe them as exact
conformal field theories. One can hope to obtain a solution
with two dimensional space-time supersymmetry  which will ensure the
vanishing of the cosmological constant. However this is expected to
keep the leading order solution unchanged even at higher orders. A
more likely scenario for obtaining the CFT for our solution may be
the construction of an
exact string action with underlying $N=2$ local worldsheet
supersymmetry. For such a theory, in target space
dimension $D=2$, the tree level cosmological constant vanishes.
Along with these aspects, we are also planning to present
a large class of solutions based on the techniques of this paper
in a forthcoming publication.

\vfil
\eject
\end{document}